\documentclass[amssymb,prb,twocolumn]{revtex4}
\usepackage{graphicx}
\newfont{\bl}{cmbxsl10 scaled\magstep1}
\begin{document}
\draft
\title{Spin interactions of interstitial Mn ions in ferromagnetic GaMnAs}
\author{\framebox{J. Blinowski}}
\affiliation{ Institute of Theoretical Physics, Warsaw University, ul.~Ho\.za 69, 00-681 Warszawa, 
Poland}
\author{P. Kacman}
\affiliation{Institute of Physics, Polish Academy of Sciences, al.~Lotnik\'ow 32/46, 
02-668 Warszawa, Poland}
\date{\today}
\begin{abstract}
The recently reported Rutherford backscattering and particle-induced X-ray emission experiments
\cite{Yu-prb02} have revealed 
that in low-temperature MBE grown Ga$_{1-x}$Mn$_{x}$As a significant part of the incorporated 
Mn atoms occupies tetrahedral interstitial sites in the lattice. 
Here we study the magnetic properties of these interstitial (Mn$_{\text{I}}$)  ions. We show 
that they do not participate in the 
hole-induced ferromagnetism. Moreover, Mn$_{\text{I}}$ double donors  may form pairs with 
the nearest  substitutional (Mn$_{\text{Ga}}$) acceptors - our calculations evidence that  
the spins in such pairs are antiferromagnetically coupled by the superexchange. 
We also show that for the Mn ion 
in the other, hexagonal, interstitial position (which seems to be the case in the 
Ga$_{1-x-y}$Mn$_{x}$Be$_{y}$As samples)
the p-d interactions with the holes, responsible 
for the ferromagnetism, are very much suppressed.
\end{abstract}

\maketitle

The incorporation of transition metal ions into the III-V host 
semiconductors by low-temperature molecular beam epitaxy (LT MBE), 
i.e., the discovery of ferromagnetic dilute magnetic semiconductors (DMS) in
the pioneering work by Munekata {\it et al.}\cite{Munekata}, 
was a major step towards the integration of the spin degrees of freedom 
with the semiconducting properties in the same material. Still, 
the prospects for practical applications  of  DMS
in "spintronic" devices depend crucially on the possibilities to increase in these materials the
 temperature of the transition to the ferromagnetic phase. 
The highest Curie temperatures (T$_{\text{C}}$) in DMS 
have been obtained by a substitution of Mn for Ga in 
GaAs, which was complemented by a post-growth annealing in temperatures only slightly exceeding 
the LT MBE growth temperature. 
Until  recently the T$_{\text{C}}$=110~K seemed to be the upper limit
for this material \cite{Ohno99, Potashnik, Hayashi, Ishiwata}. In the last months, however, considerably 
higher values of  
T$_{\text{C}}$ in GaMnAs, even exceeding 150~K for thin films, have been reported by several 
groups\cite{Kuryliszyn, 
Edmonds, Ohno02, Samarth02}. This progress has been made basically by an optimization of the 
annealing time and temperature. 
 
In the theoretical models describing the ferromagnetism in DMS (e.g., in Ref.~\onlinecite{Dietl, Konig, Sanvito}) 
the T$_{\text{C}}$  is expected to increase with both, the magnetic ions and hole concentrations. 
In LT MBE grown Ga$_{\text{1-x}}$Mn$_{\text{x}}$As this was indeed experimentally 
established for Mn concentrations up to about x=0.07, \cite{Ohno99}. 
The Mn ion in the substitutional position in the GaAs lattice (Mn$_{\text{Ga}}$) acts as an 
acceptor, but in 
all Ga$_{\text{1-x}}$Mn$_{\text{x}}$As samples the hole concentration is substantially lower 
than the Mn content. 
This has been ascribed to the presence of compensating donors, in particular to the formation 
of arsenic antisites 
(As$_{\text{Ga}}$) during the epitaxial growth of Ga$_{\text{1-x}}$Mn$_{\text{x}}$As at As 
overpressure \cite{Sanvito1, Sadowski}. 
In Ref.\onlinecite{Potashnik,  Ishiwata} and \onlinecite{Korzhavyi}  the observed 
annealing-induced changes 
of the T$_{\text{C}}$ were attributed solely to the decrease of the concentration of arsenic 
antisites leading to the 
increase of the hole concentration. These antisites, however, are relatively stable defects - 
it was shown that to remove 
As$_{\text{Ga}}$ from LT MBE grown GaAs the annealing temperatures above 450~C are 
needed \cite{Bliss}. Recently, simultaneous channeling Rutherford backscattering (c-RBS) 
and particle-induced X-ray emission (c-PIXE) 
experiments shed new light on this problem \cite{Yu-prb02}. Namely, they have revealed that 
in LT MBE grown ferromagnetic Ga$_{\text{1-x}}$Mn$_{\text{x}}$As with high x
a significant fraction
of incorporated Mn atoms (ca 15\% for the as-grown Ga$_{\text{0.91}}$Mn$_{\text{0.09}}$As 
sample) occupies 
well defined, commensurate with the GaAs lattice interstitial positions.

\begin{figure}
\includegraphics*[width=85mm]{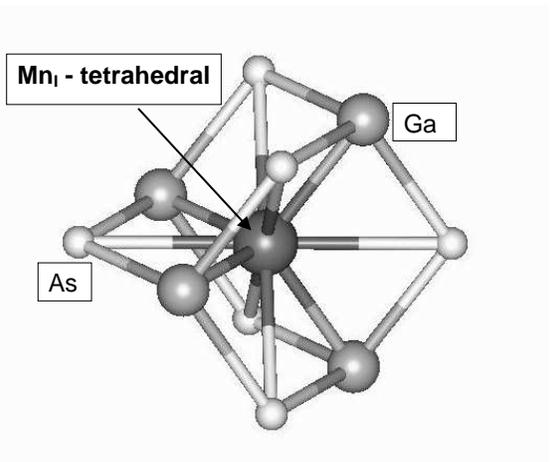}
\caption[]{The  nearest four  cation  and six  anion neighbors for an ion in the 
tetrahedral interstitial position in the zinc-blende lattice.}
\label{Mnt}
\end{figure}

In the diamond cubic crystal lattice there are two possible interstitial positions, the so called
tetrahedral and hexagonal sites, in which the atoms are shadowed along $\langle 100 \rangle$ and
$\langle 111 \rangle$ direction and exposed in the $\langle 110 \rangle$ axial channel, as seen at the experiment. They can be 
distinguished by studying angular scans around the $\langle 110 \rangle$ axial direction \cite{Feldman}.  The
scans presented in Ref.~ \onlinecite{Yu-prb02} suggested that the 
interstitial Mn ions (Mn$_{\text{I}}$) observed in Ga$_{\text{1-x}}$Mn$_{\text{x}}$As occupy the
tetrahedral sites, in which the interstitial is surrounded by four nearest neighbors, as presented 
in  Fig.~\ref{Mnt}.

 The Mn$_{\text{I}}$ serve, 
like As$_{\text{Ga}}$, as double donors, decreasing the hole concentration. 
The results presented 
in Ref.~\onlinecite{Yu-prb02} directly showed that in the 
process of LT annealing the 
Mn$_{\text{I}}$ ions are moved to random, incommensurate with the GaAs lattice  positions 
(e.g., MnAs clusters),   in which the Mn ions are electrically inactive.
Thus, in the annealed samples the concentration of the compensating Mn$_{\text{I}}$ donors 
decreases considerably whereas
the hole concentration increases and the observed Curie temperature is much higher. 
Moreover, it was demonstrated that the appropriate annealing 
increases the saturation magnetization, i.e., that the presence of Mn$_{\text{I}}$ reduces 
the net magnetic moment\cite{Yu-prb02,Yu-apl02,Kuryliszyn, JB02}.

The described above experimental results stimulated theoretical studies on the formation and
properties of interstitial Mn in the GaMnAs ternary compound. First, the electronic structure of
the GaMnAs with Mn in substitutional and interstitial position was calculated by {\it ab initio} methods,
showing that indeed Mn interstitials act as double donors \cite{Maca}. In a recent paper
\cite{Erwin} the self-compensation of Mn in such semiconductors was studied within the density-functional theory. In Ref.~\onlinecite{Erwin} it was shown that interstitial Mn can be easily formed near the surface.   
Here we consider the spin properties of interstitial magnetic ions.
 We study the spin interactions for the Mn$_{\text{I}}$ ion in the 
tetraheadral interstitial position in order to provide 
theoretical basis for the understanding of the experimental findings 
concerning the magnetic behavior of the as-grown and annealed 
Ga$_{\text{1-x}}$Mn$_{\text{x}}$As samples. We analyze the hybridization 
of the d-orbitals of these ions with the valence band p-states.  This effect is essential 
for both, the superexchange and the RKKY-type, dominant ion-ion 
interactions in DMS.  
It is widely accepted that the latter mechanism is responsible for the hole-induced 
ferromagnetism in III-V DMS and that the T$_{\text{C}}$ depends crucially on the 
p-d hybridization - within the Zener model \cite{Dietl} T$_{\text{C}}$ is proportional to 
the square of the 
kinetic p-d exchange constant $\beta$, i.e., to the fourth power of the 
hybridization constant {\it V} at the centre of the Brillouin zone.

The valence band states in Ga$_{\text{1-x}}$Mn$_{\text{x}}$As are built primarily from the 
anion p-orbitals, 
thus the p-d hybridization for a given magnetic ion is determined by the positions 
of its nearest-neighbor anions.  In zinc-blende 
lattice of GaAs, the Mn ion in the cation substitutional position has four 
anion nearest neighbors at the distances $a\sqrt{3}/4$  (where  $a$ is the lattice constant)
along the [1, 1, 1], [1,-1,-1], [-1, 1,-1] and 
[-1,-1, 1] directions. For these positions the inter-atomic matrix elements, $E_{x,xy}$, 
$E_{x,yz}$, $E_{x,zx}$, etc., expressed in terms of the Harrison parameters $V_{pd\sigma}$ and $V_{pd\pi}$ \cite{Harrison}, add up constructively to the hybridization constant {\it V} in the hybridization 
Hamiltionian $\hat H_h$, with different weights for 
different points of the Brillouin zone. At the point $\vec{k}$=0 of the Brillouin zone they sum up to the value:
$4(V_{pd\sigma}-2/\sqrt{3}V_{pd\pi})$.

\begin{figure}
\includegraphics*[width=85mm]{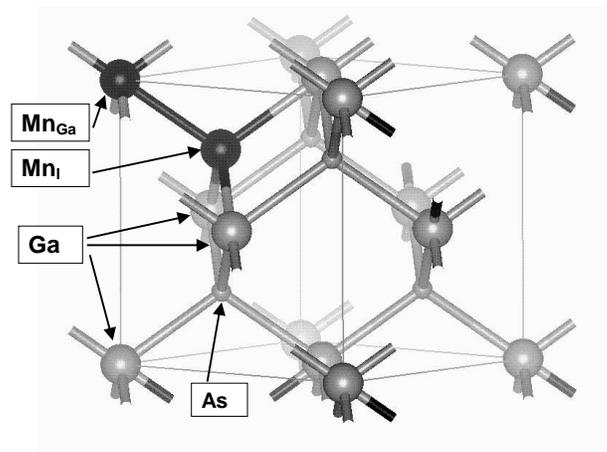}
\caption[]{Mn$_{\text{Ga}}-$Mn$_{\text{I}}$ pair in the GaAs structure.}
\label{pair}
\end{figure}

In contrast, the ion in a tetrahedral interstitial position,
 e.g., ($\frac{1}{4}\frac {1}{4}\frac {3}{4}$) as in 
Fig.~\ref{pair}, has 6 anion neighbors on [0, 0, $\pm$ 1], [0, $\pm$ 1, 0] and [$\pm$1, 0, 0] 
directions, 
at the distances $a/2$ (see Fig.~\ref{Mnt}).  In this case all not equal zero inter-atomic 
matrix elements are proportional to the appropriate $\sin{(ak_{i}/2)}$ (where $k_i$, 
$i=x,y,z$, are the 
components of the wave vector $\vec{k}$) and they vanish at the centre of the
Brillouin zone.  Thus, the Mn$_{\text{I}}$ d-orbitals do not hybridize 
with the p-states of the holes at the top of the valence band 
(i.e. for the tetrahedral interstitials the kinetic exchange constant $\beta_{I_t}$= 0) and they do not 
contribute 
to the hole-induced ferromagnetism.  This means that the formation of Mn interstitials  
decreases not only the hole concentration but also the number of Mn ions participating 
in the Zener-type ferromagnetism.  Still, these effects do not explain why the removal of 
interstitials leads to the increase of magnetization and to the higher 
T$_{\text{C}}$ than expected from the rise of the hole concentration \cite{Yu-apl02}.  

As pointed out already by Yu {\it et al.}\cite{Yu-prb02} the electrostatic 
attraction between positively charged Mn$_{\text{I}}$  donors 
and negative Mn$_{\text{Ga}}$  acceptors stabilizes the otherwise highly mobile 
Mn$_{\text{I}}$ in the interstitial sites adjacent to Mn$_{\text{Ga}}$, forming
a Mn$_{\text{Ga}}$-Mn$_{\text{I}}$ pair, as shown in Fig.~\ref{pair}. 
One notices that despite the fact that for the interstitials the p-d kinetic exchange 
and, consequently, the hybridization mediated spin interactions with holes in the vicinity
of the top of the valence band vanish, the ionic spins in the pair
can be coupled by superexchange mechanism.
In the latter process the spins of the two ions, $\vec{S}_1$ and $\vec{S}_2$, are correlated due to 
the spin-dependent 
p-d exchange interaction between each of the ions and the valence band electrons 
in the entire Brillouin zone. The superexchange Hamiltonian:
\begin{equation}
\hat{H}_{superexchange}=-2J(\vec{R}_{12})\hat{\vec{S}}_1\cdot\hat{\vec{S}}_2
\end{equation}
can be obtained by a proper selection of spin-dependent terms in the matrix of the 
fourth order perturbation with respect to the hybridization for a system of two ions 
in the crystal:
\begin{equation}
-\sum_{l, l', l''}\frac{\langle f\mid \hat H_{h}\mid l''\rangle
\langle l''\mid \hat H_{h}\mid l'\rangle\langle l'\mid \hat H_{h}\mid l\rangle
\langle l\mid \hat H_{h}\mid i\rangle}{(E_{l''}-E_0)(E_{l'}-E_0)(E_l-E_0)}
\end{equation}

  Using the virtual transition picture, one can say that the superexchange 
is a result of four virtual transitions of an electron - from the band onto the d-shell of the ion and from 
the  ionic d-shell to the band, in different sequences \cite{Kacman}.  The quantitative determination of 
the superexchange constant $J$ requires the knowledge of the energies of these virtual transitions, which
are represented by the energy differences between the intermediate and initial states of the system of two ions and the  completely filled valence bands, in
the denominator of Equation (2). Of primary importance it is, however, to determine the sign of 
the superexchange interaction for the Mn$_{\text{Ga}}$-Mn$_{\text{I}}$ pair. In the following, 
we calculate the exchange constant $J$ within 
a simplified model, in which we neglect the dispersion of the valence bands but we account 
for the wave-vector dependence of the hybridization matrix elements.  
The resulting formula for the exchange constant $J$ reads:
\begin{widetext}
\begin{center}
\begin{eqnarray}
 J( \vec{R}_{12})= -\frac{1}{25}\Biggl[ \frac{1}{E_{a_1}E_{a_2}}\biggl(\frac{1}{E_{a_1}}+\frac{1}{E_{a_2}}\biggr)+
\frac{1}{E_{a_1}^2(E_{a_1}+E_{d_2})}+\frac{1}{E_{a_2}^2(E_{a_2}+E_{d_1})}\Biggr] \times \\ \nonumber 
\times \sum_{\nu_1,\nu_2,\vec{k}_1,\vec{k}_2,m,n}V_{\nu_1,\vec{k}_1,m}^{*}(2)V_{\nu_2,\vec{k}_2,m}(2)
V_{\nu_2,\vec{k}_2,n}^{*}(1)V_{\nu_1,\vec{k}_1,n}(1)
\end{eqnarray}
\end{center}
\end{widetext}

In Equation (3) the summation runs over the valence band indices $\nu_1$ and $\nu_2$, 
the wave-vectors $\vec{k}_1$, $\vec{k}_2$  from the entire Brillouin zone, and over the Mn d-orbitals 
$m$, $n$. 
The energies $E_{a_i}$  and $E_{d_i}$ ({\it i} = 1,2) are the transfer energies for the 
electron from the valence 
band onto the ion {\it i} ("acceptor") and from the ion {\it i} to the valence band ("donor"), respectively. 
It should be noted
that these energies for the interstitial Mn ion are completely unknown.   Still, since all these 
energies as well as the sum, which we calculated numerically, are positive, we can conclude 
that the Mn$_{\text{Ga}}$-Mn$_{\text{I}}$ pair is {\it antiferromagnetically} coupled. 
Thus, Mn ions
when in tetrahedral interstitial positions not only do not contribute 
to the hole-induced ferromagnetism but they also make some of the substitutional Mn ions 
magnetically inactive by forming with them close pairs, in which the spins of the ions are
antiferromagnetically coupled by the superexchange mechanism. This explains 
the experimental observations that the removal of  Mn$_{\text{I}}$ ions by low-temperature 
annealing leads not only to an increase of the hole concentration, but also to a significant 
increase of the magnetization.  

To estimate the strength of this coupling we compare $J$ with the superexchange 
constant $J'$ for a Mn$_{\text{Ga}}$-Mn$_{\text{Ga}}$ closest pair, obtained within the same 
simple model.  
Using the same transition energies for both Mn$_{\text{Ga}}$ and Mn$_{\text{I}}$ ions, 
we obtain $J/J'\approx 1.6$.  
This is not surprising in view of the small distance between the interstitial and the 
nearest substitutional Mn ions and the larger number of anion neighbors for Mn$_{\text{I}}$.  
With a reasonable value of 3~eV for the Mn$_{\text{Ga}}$ charge transfer energies, with the values of 
Harrison parameters $V_{pd\sigma}$=1.1~eV \cite{Okabayashi}, and $V_{pd\pi}=-\frac{1}{2}V_{pd\sigma}$,
(which for the Mn$_{\text{I}}$ we scale according to the Harrison's prescription) \cite{Harrison} 
the absolute values of $J$ and $J'$ 
constants are by far not negligible: $J\approx$ 71~K and $J'\approx$ 43~K.

\begin{figure}
\includegraphics*[width=85mm]{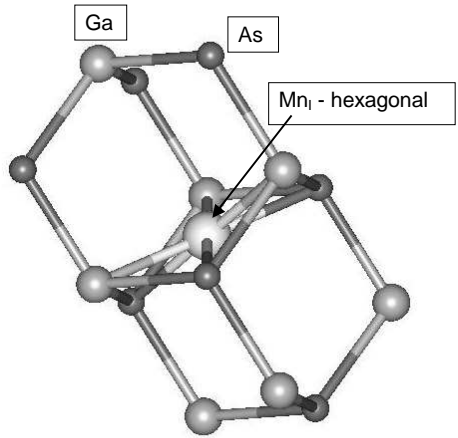}
\caption[]{The six  (three cations and three anions) nearest neighbors and the next four cations  and four 
anions for an ion in the hexagonal interstitial position in the zinc blende lattice.}
\label{Mnh}
\end{figure}

The role of interstitial Mn ions occurred to be even more pronounced in Ga$_{1-x-y}$Mn$_{x}$Be$_{y}$As 
samples, grown at Notre Dame with the hope to increase the hole concentration, and hence T$_{\text{C}}$, 
by introducing another acceptor \cite{Wojtowicz}. Instead, it turned out that adding Be to Ga$_{1-x}$Mn$_{x}$As 
increases the concentration of Mn$_{\text{I}}$ at the expense of Mn$_{\text{Ga}}$\cite{JB02, ICPS02}.  
At the same time, although the hole concentration 
does not change significantly, the T$_{\text{C}}$ drops dramatically \cite{Wojtowicz, Wojtowicz1}, in agreement with
the presented above result that the Mn$_{\text{I}}$ do not participate in the hole-induced ferromagnetism.
Recently performed angular scans seem to suggest, however, that the Be$_{\text{Ga}}$ acceptor
stabilizes the Mn interstitial donor not in the tetrahedral but in the hexagonal, 
( $\frac{3}{8}\frac{5}{8}\frac{3}{8}$ ) position\cite{preprint}. In this site the Mn$_{\text{I}}$ 
has three anion nearest neighbors, as shown in Fig.~\ref{Mnh}, on the [-1, -3, -1], 
[3, 1, -1] and [-1, 1, 3] directions, at the distance $a\sqrt{11}/8$. In such case one does not
expect the kinetic p-d exchange constant $\beta$ to be equal to zero - it can be rather expected
that the hybridization for the ion in this site should be stronger,  due to the smaller distance to the anions, what increases the Harrison parameters 
$V_{pd\sigma}$ and $V_{pd\pi})$. Surprisingly enough, the
most of the inter-atomic matrix elements in the hybridization constant $V$ for the hexagonal interstitial Mn mutually cancel at the centre of the Brillouin zone. This leads to a considerably smaller than for the substitutional Mn ion value of the kinetic exchange constant $\beta_{I_h}$, i.e.,  
$\beta/\beta_{I_h} \approx 5$.  As the Curie temperature in the Zener model 
depends on $\beta^2$, we conclude that the contribution to the hole-induced ferromagnetism  from the
Mn ions occupying the hexagonal interstitial sites is as well very much suppressed.
  
In conclusion, we have shown that not only the compensating properties of the interstitial magnetic ions impose a limit to the Curie temperature  in the ferromagnetic Ga$_{1-x}$Mn$_{x}$As and Ga$_{1-x-y}$Mn$_{x}$Be$_{y}$As samples.  Also their magnetic properties in both (tetrahedral and hexagonal)  interstitial sites, i.e., the negligible kinetic exchange constant and strong antiferromagnetic superexchange with the adjacent  substitutional Mn ion, act towards diminishing the transition temperature. 
\begin{acknowledgments}
The authors are very much obliged to J. Furdyna, T. Wojtowicz and W. Walukiewicz for making their unpublished results  available to us and for  valuable discussions and comments. Support of the FENIKS project (EC:G5RD-CT-2001-00535) and of the Polish State Committee for Scientific Research grant PBZ-KBN-044/P03/2001 is also gratefully acknowledged.
\end{acknowledgments}

\end{document}